\documentclass[amsmath,amssymb, aps, prb,reprint,superscriptaddress]{revtex4-1}

\usepackage{graphicx}

\def\bd{\begin{displaymath}}
\def\be{\begin{equation}}
\def\ed{\end{displaymath}}
\def\ee{\end{equation}}
\def\bsub{\begin{subequations}}
\def\esub{\end{subequations}}

\newcommand{\Eq}[1]{Eq.~(\ref{#1})}
\newcommand{\Fig}[1]{Fig.~\ref{#1}}
\newcommand{\abs}[1]{\left| #1 \right|}

\makeatletter
\newcommand*\dashline{\rotatebox[origin=c]{90}{$\dabar@\dabar@\dabar@$}}
\makeatother

\begin{document}

\title{Gaffnian holonomy through the coherent state method}

\author{John Flavin}

\affiliation{
Department of Physics and Center for Materials Innovation, 
Washington University, St. Louis, MO 63130, USA}

\author{Ronny Thomale}

\affiliation{Department of Physics, Stanford University, Stanford, CA 94305, USA}

\author{Alexander Seidel}

\affiliation{
Department of Physics and Center for Materials Innovation,
Washington University, St. Louis, MO 63130, USA}

\begin{abstract}
	We analyze the effect of exchanging quasiholes described by Gaffnian quantum Hall trial state wave functions. 
	This exchange is carried out via adiabatic transport using the
        recently developed coherent state Ansatz.  
        We argue that our Ansatz is justified if the Gaffnian parent Hamiltonian has a charge gap, 
	even though it is gapless to neutral excitations, and may therefore properly describe 
	the adiabatic transport of Gaffnian quasiholes.
	For nonunitary states such as the Gaffnian, the result of adiabatic transport cannot agree with the 
	monodromies of the conformal block wave functions, and may or may not lead to 
	well-defined anyon statistics. Using the coherent state Ansatz, we find two unitary solutions for the statistics, 
	one of which agrees with the statistics of the non-Abelian spin-singlet state by Ardonne and Schoutens.
\end{abstract}


\maketitle


\section{Introduction\label{intro}}

The study of fractional quantum Hall (FQH) liquids has lead to the discovery of rich classes
of topological phases. The rapid progress in the theoretical exploration of this particular niche
of correlated electron systems was made possible by the development of a set of principles for the 
construction of ``fixed point'' or ``prototype'' wave functions for possible FQH phases,
as pioneered by Laughlin\cite{laughlin} and then greatly expanded by other seminal 
contributions.\cite{halperin, MR, HR, RR}
This progress was compounded by the discovery of a link between these special wave functions
and conformal field theory (CFT).\cite{MR} Via this link the topological quantum field theory 
describing a state can be inferred, which for some cases of interest is non-Abelian.\cite{MR}
The possibility of anyonic quasiparticles with non-Abelian braiding statistics may be
the most spectacular implication of this field theoretic mapping, and its experimental
verification remains a profound challenge to date, with topological quantum computing
being a proposed application.\cite{kitaev, freedman}
Even in theory, a direct verification of the implied field theoretic mapping is nontrivial,
and involves the calculation of adiabatic transport of quasiparticles using full many-body
wave functions. This is still difficult in general, but has been possible for Abelian states\cite{ASW}
and recently for $p+ip$ wave superfluids and Moore-Read-type quantum Hall states.\cite{read09, read_preprint,bonderson_gurarie_nayak}
The general interpretation of analytic trial wave functions as describing topological phases
is thus well accepted whenever these wave functions are related to the conformal blocks of a unitary
rational CFT. In this case, the field theoretic mapping described above implies that the statistics
of quasihole-type excitations, as defined through adiabatic transport, 
is directly encoded in the monodromies of the associated conformal
blocks.\cite{MR, nayakwilczek} This means that for such conformal block wave functions, adiabatic transport
is the same as analytic continuation in the quasihole coordinates, and Berry phase effects are trivial.
This then allows one to obtain the quasihole statistics directly from wave functions, but without explicitly calculating
the Berry connection, or Wilczek-Zee connection, that defines adiabatic transport.\cite{Berry,WZ} Indeed, in may cases of interest, 
this remains a profound technical challenge.

In addition to wave functions arising from unitary CFTs, there is also considerable interest in 
analytic trial states that are similarly related to nonunitary CFTs.\cite{HR,gaffnian}
The physical interpretation of such states remains much more subtle.
Here, the field theoretic mapping employed in the unitary case does not lead to a topological 
quantum field theory that can serve as the low energy effective theory of the state in question.
In particular, the conformal block monodromies cannot be interpreted to describe adiabatic transport,
as they do not result in unitary transformations on states. In contrast, adiabatic transport describes
(a limit of) the time evolution governed by a Hermitian Hamiltonian, and is therefore always described
by a unitary transformation. For these reasons, it has been 
argued\cite{gaffnian, read09, read_preprint, read09_2, readgreen}
that states obtained from nonunitary CFTs describe gapless critical points within the phase diagram of
quantum Hall states, especially in those cases where a local parent Hamiltonian for the state exists.
Examples of the latter kind include the Haldane-Rezayi (HR) state,\cite{HR} and the state known as Gaffnian,
which has surfaced in the literature as early 1993 through a fixed many-body clustering property,\cite{greiter}
and which has been thoroughly discussed and was proposed to be critical using a CFT construction.\cite{gaffnian}

The question arises what hidden orders can be identified in such nonunitary states, 
whether they be remnants of topological orders or orders of a different kind.
Unfortunately, there is currently no efficient and universally applicable method to test for the topologically ordered\cite{wenniu}
nature of a state directly through the study of ground state properties. 
Much progress along these lines has recently been made through
the analysis of entanglement spectra,\cite{li08,thomale} which are directly related to edge spectra. 
It has been argued that the edge spectrum of the Gaffnian is inconsistent with that of any unitary CFT, 
and that this contradicts the existence of a gap in the bulk spectrum,\cite{read09_2}
which is required for a topological phase. 
In principle, topological orders can be detected through nonlocal order parameters,\cite{readorder}
though it remains difficult to explicitly construct such objects for general non-Abelian topological orders in microscopic quantum Hall wave functions.
The situation is similar in quantum magnetism. There, nonlocal operators detecting a topological phase can be directly constructed for toy models\cite{misguich} 
defined on highly constrained Hilbert spaces where a gauge structure is explicit
(see Ref. \onlinecite{gregoretal} for a general discussion). However, such order parameters generally remain elusive in models where similar physics is emergent 
within the low energy sector of a larger Hilbert space (e.g., Ref. \onlinecite{seidelkagome}). The situation is much simpler in one-dimensional systems exhibiting a 
Haldane or Luttinger liquid phase. The hidden orders of these phases can be probed through nonlocal objects measuring squeezed particle configurations,\cite{RSOS,kruis}
and their origin is quite manifest, e.g., in certain limits of Luttinger liquids where the wave function assumes a special factorized form.\cite{ogata90,ogata91,seideltJJ,Ribeiro2006}
For topological orders, on the other hand, the most direct probe that can, in principle, be implemented at a microscopic wave function level is the study
of the braiding statistics of localized elementary excitations.

In this paper, we are interested in some formal properties of the Gaffnian state.
We ask the question whether the Gaffnian trial wave functions may define
some unitary anyon model through the holonomy calculated along exchange paths.
Somewhat more physically speaking, this corresponds the the adiabatic transport of trial state quasiholes 
in the presence of a finite size gap. Indeed, this question is mathematically well defined. The quasihole 
trial states can be characterized as the unique zero-energy eigenstates (zero modes) 
of a local parent Hamiltonian.\cite{greiter,gaffnian} For given quasihole configuration, the associated 
conformal block wave functions define a finite-dimensional subspace, which can be interpreted
as a fiber over a point in the quasihole configuration space. The question is thus whether the holomomy
associated with exchange paths in this configuration space induces well-defined statistics.
It is clear from the outset that if this is so,
the holonomies must be quite different from the conformal block monodromies,
since these holonomies give rise to unitary transformations on fibers.
Physically, this is clear from the fact that these holonomies describe the adiabatic
transport of quasiholes protected by a finite size gap. Mathematically,
it follows from the fact that the connection on our vector bundle is a Wilczek-Zee connection 
defined in terms of a physical scalar product.

The question defined above can be rigorously addressed only by calculating
the Wilczek-Zee connection from the given analytic wave functions. Unfortunately,
we do not know how to do this for the Gaffnian state. Instead, we will use this question as
a testbed for a recently developed coherent state 
method\cite{seidel_lee, seidel_pfaffian, flavin11} to calculate adiabatic transport 
of quasiholes based on the ``thin torus'' (TT) \cite{seidel1,
karlhede1, karlhede2, karlhede3, seidel_lee, karlhede4, seidelyang,
seidel_pfaffian, karlhede5, karlhede09, seidel_sduality, seidelHR, flavin11} or 
``dominance''\cite{greiter, haldanebernevig,haldanebernevig3, Bernevig2008}
patterns of the wave function. This method has been shown to be quite efficient 
for a number of interesting states based on unitary CFTs, but has thus far not been applied
to the nonunitary case. Our motivation to clarify  the applicability of this method to a nonunitary
state is twofold. A negative result (no consistent anyon model) would further strengthen the 
case that the TT limit contains information about the gapped or gapless nature of the underlying
state. This has been explored by one of us recently for the HR state,\cite{seidelHR}
though not with regard to statistics. On the other hand, if a consistent anyon model
is obtained, we can argue that this is at least a very plausible scenario for the
holonomies defined by the Gaffnian quasihole states, as we will further elaborate below.

The coherent state method is based on the assumption of adiabatic continuity
between states defined on a torus with arbitrary aspect ratio and corresponding states
in the thin torus limit. It further rests on general assumptions about a coherent state
Ansatz for localized quasi holes in terms of adiabatically continued TT states.
Detailed arguments in favor of this Ansatz have been given in Ref. \onlinecite{flavin11}.
Some of these arguments, in particular the justification for the factorized form of the
Ansatz (see \Eq{phi} below), also rest on a notion of locality, which is more subtle 
in a gapless state. We argue however, that the necessary assumptions still apply,
as long as there is a finite gap in the {\em charge} sector of the system,
independent of the existence of gapless neutral excitations.
The scaling of the charge gap of the Gaffnian state
has been discussed in some detail in Ref. \onlinecite{toke}, but at the moment, the 
question whether it remains finite in the thermodynamic limit has not been conclusively
resolved to the best of our knowledge.

There is a close connection between TT patterns and CFT fusion rules, which has been
elaborated in Refs. \onlinecite{ardonne_gaffnian,ardonne2008}.
In terms of the data used to construct an anyon model, there is thus some
similarity between the present method, and the procedure
of constructing $F$-matrices consistent with given fusion rules,
using the axioms of modular tensor categories.\cite{bonderson_thesis, bonderson_interferometry}
Indeed, the two methods have so far given consistent, if not always identical, results in 
the unitary cases.\cite{flavin11} Some differences between these two approaches are worth noting.
While the connection with tensor category theory can be physically justified
using the general framework of local quantum field theory,\cite{froehlich}
the assumptions used in the coherent state approach are not field theoretic in character.
Indeed, the $F$-matrix, although it could be ultimately constructed, does not directly appear in this approach,
and none of the defining consistency equations of this approach make reference to it.
Moreover, the coherent state method, being ultimately based on adiabatic transport,
could not in principle lead to a nonunitary anyon model in its current formulation.
A secondary, but non-negligible motivation for this work is thus to shine further light
onto the connection between these two different methods in the context of a nonunitary state.

The paper is organized as follows. In Sec. \ref{braid} we will summarize our calculation,
which leads to two closely related anyon models of Fibonacci type. All details are
relegated to two Appendices. In Sec. \ref{discussion}, we discuss our results,
and make contact with the non-Abelian spin singlet (NASS) of Ref. \onlinecite{ardonne3}.
In a future version of this preprint soon to follow, we will also present new numerical studies on the
zero mode spectrum of the Gaffnian state and its connection to root partitions.  

\section{Braiding statistics through the coherent state method\label{braid}}

Our method to extract the statistics of a state from thin torus patterns, based on the assumptions discussed in the Sec. \ref{intro}, 
has been documented in detail in a recent paper.\cite{flavin11} Here we will describe its setup for the Gaffnian state, and focus on 
differences that arise compared to the discussion of the ($k=3$) Read-Rezayi state\cite{RR} as given in Ref. \onlinecite{flavin11}.

The thin torus patterns of the bosonic $\nu=2/3$ Gaffnian, and their relation to the underlying minimal model CFT,  have been thoroughly discussed
by Ardonne.\cite{ardonne_gaffnian} These patterns can be identified in the usual 
way\cite{seidel1,karlhede1, karlhede2, karlhede3, seidel_lee, karlhede4, seidelyang,
seidel_pfaffian, karlhede5, karlhede09, seidel_sduality, seidelHR, flavin11} when wave functions on the torus are considered
in the formal limit of an ``extreme aspect ratio'', e.g. $L_x\gg1,L_y\ll1$. (Here and in the following, we set the magnetic length $\ell_B$ equal to $1$.)
The patterns emerging in this limit are occupancy numbers in a suitably chosen Landau level basis on the torus (see, e.g., \onlinecite{flavin11}),
describing the trivial non-interacting product state resulting in this limit. For the six degenerate Gaffnian ground states, these patterns read
$200200\dots$, $020020\dots$, and $002002\dots$, which we will call ``(200)-type'', or $011011\dots$, $101101\dots$, and $110110\dots$, which we
refer to as ``(011)-type''.  

As usual,\cite{seidel1, seidel_lee, seidelyang, seidel_pfaffian, seidel_sduality, seidelHR, flavin11, 
karlhede1, karlhede2, karlhede3, karlhede4, karlhede5, karlhede09} we will assume that states defined on a 
``bulk torus''---$L_x\gg 1,L_y\gg 1$---can be evolved adiabatically into the thin torus limit described
above, where the evolution is governed by the Gaffnian parent Hamiltonian\cite{gaffnian} with slowly varying aspect ratio.
Likewise, states with quasiholes on the bulk torus are assumed to evolve into elementary domain walls between ground state patterns. There are three elementary domain wall strings: 100 and 001, which occur between (200)-type and (011)-type ground states, and 010, which occurs between two different (011)-type ground states. These domain walls may link various different combinations of ground state patterns, thus forming charge $1/3$ solitons.
Representative examples are given by the strings $\dotsc01101\mathbf{100}200200\dotsc$ and $\dotsc011011\mathbf{010}110110\dotsc$. 
All charge $1/3$ domain wall patterns occurring in the TT limit are locally given by translated and/or inverted versions of these two types of
strings. This establishes a notion of ``fusion rules'' in the TT limit.\cite{ardonne_gaffnian}

The thin torus states are naturally labeled by the positions of the domain walls together with a topological sector label---i.e.,
a label identifying the sequence of patterns between the domain walls. We use the notation
 $|a_1,\dots,a_n;c,\alpha)$ for the TT states appearing in the limit $L_x\gg1,L_y\ll1$, and
$\overline{|a_1,\dots,a_n;c,\alpha)}$ for the states appearing in the $L_x\ll 1,L_y\gg 1$ limit. 
The latter are simple product states in a ``rotated'' or dual Landau level basis\cite{seidel_lee, seidel_pfaffian, flavin11},
and must be well distinguished from the states  $|a_1,\dots,a_n;c,\alpha)$. Here,
the $a_i$ label the orbital positions of the domain walls.
For the topological
sector labels, we follow the convention of \onlinecite{flavin11}, where $\alpha$ distinguishes 
classes of sectors that are not related by translation, and for given $\alpha$, $c=-1,0,1$
distinguishes the three translationally related members of each class. Below we will also
use translational properties to define a unique convention for how the $c$ labels are to be 
assigned. As usual\cite{flavin11}, the domain wall positions can change only
by multiples of a certain ``stride'' within each topological sector (here, multiples of $3$),
and are thus of the general form $a_j = 3 n_j+f_j(c,\alpha)$. $f_j(c,\alpha)$ is 
an offset factor that depends on the sequence of domain walls (i.e., the topological sector and the domain wall in question).
For symmetric domain walls, its value is uniquely determined by symmetry. However, for asymmetric domain
walls, a certain ambiguity exists a priori in how to define the domain wall position precisely with respect
to the adjacent orbitals. This is accounted for by the shift---or asymmetry---parameter $s$. See \onlinecite{flavin11}
and table \ref{table_2sectors} for details.

\begin{table}[!tbp]  
   \begin{tabular}{ccc | c || c | c }
	& $\alpha$ && Thin torus pattern & $f_1(\alpha)$ & $f_2(\alpha)$ \\
	\hline
	& 1 && $002002\mathbf{00\underline{1}}101101\mathbf{10\underline{0}}2002002$ & $-s$ & $-2+s$ \\
	& 2 && $1101101\mathbf{1\underline{0}0}2002002\mathbf{\underline{0}01}10110$ & $-1+s$ & $2-s$ \\
	& 3 && $11011001\mathbf{\underline{0}10}11011\mathbf{0\underline{1}0}110110$ & $1$ & $0$ 
   \end{tabular} 
   \caption{$c=0$ thin torus patterns for a two-quasi-hole Gaffnian state, 
   	and the offset functions of the associated domain walls. The elementary domain wall strings are in 
	bold, and the orbital positions, $3n_j$, are underlined. 
   	Patterns for $c=1(-1)$ can be obtained by shifting each occupancy number one orbital to the
	right (left), and the shift functions obtained using $f_j(c,\alpha)=f_j(\alpha)+c$.} 
   \label{table_2sectors} 
\end{table}

By the assumption of adiabatic continuity, the ``bare'' domain wall states, ${|a_1,\dots,a_n;c,\alpha)}$ and $\overline{|a_1,\dots,a_n;c,\alpha)}$,
give rise to two different complete sets of zero energy states (zero modes) of the Gaffnian parent Hamiltonian, for any aspect ratio of the torus. 
These two mutually dual zero-mode bases
are obtained by adiabatically evolving the respective bare states by means of a slow change in aspect ratio, and are denoted by
$|a_1,\dots,a_n;c,\alpha\rangle$ and $\overline{|a_1,\dots,a_n;c,\alpha\rangle}$ respectively (with the dependence on aspect ratio understood).
These types of states generally describe quasiholes that are 
localized in $x$ or $y$, respectively, at the domain wall positions, and are delocalized
in the perpendicular direction (respectively $y$ or $x$).
Their completeness within the zero-mode space implies that states with localized quasiholes (in both $x$ {\em and} $y$) can be obtained by forming proper linear
combinations. For these coherent states, with $n$ quasiholes localized at complex positions $h_j = h_{jx}+i h_{jy}$,
the following Ansatz has been motivated in previous works (see Ref. \onlinecite{flavin11}
and references therein):
\begin{align} 
	\label{psi} 
	\left| \psi_{c,\alpha}(\{h\}) \right> &= 
		{\cal N} \mspace{-18mu} \sum_{a_1<\dotsc<a_n} 
		\prod_{j=1}^n \phi_{\alpha,j}(h_j,\kappa a_j) \left| a_1,\dotsc,a_n;c,\alpha \right\rangle \\
	\label{psibar}
	\overline{\left| \psi_{c,\alpha}(\{h\}) \right>} &= 
		{\cal N'} \mspace{-18mu} \sum_{a_1<\dotsc<a_n} 
		\prod_{j=1}^n \bar\phi_{\alpha,j}(h_j,\overline\kappa a_j) 
		\overline{\left| a_1,\dotsc,a_n;c,\alpha \right\rangle} 
\end{align}
with the Gaussian amplitude form factor 
\be \label{phi}
	\phi_{\alpha,j}(h_j,a_j) = 
	\exp \left[ \frac i3 (h_{jy} + \delta(\alpha,j)/\kappa) \kappa a_j - \gamma (h_{jx} - \kappa a_j)^2 \right] \;,
\ee
and its dual counterpart $\bar\phi_{\alpha,j}(h_j,a_j) = \phi_{\alpha,j}(-i h_j, a_j)|_{\kappa \rightarrow \bar\kappa }$.
Here, $\kappa = 2\pi/L_y$, $\bar \kappa = 2\pi/L_x$, and $\delta(\alpha,j)$ can be shown\cite{flavin11} to be
$0$ or $\pi$, taking on the same value for symmetry-related (translation or inversion) domain walls.
${\cal N}$ and ${\cal N'}$ are normalization factors.

It must be emphasized that these coherent states are valid\cite{flavin11} only for well-separated quasiholes: \Eq{psi} is valid when $\abs{h_{jx}-h_{ix}}\gg 1$ for all $i,j$, and \Eq{psibar} when $\abs{h_{jy}-h_{iy}}\gg 1$ for all $i,j$. When both conditions are satisfied,
Eqs. \eqref{psi} and \eqref{psibar} describe the same zero modes for given quasihole positions, and must be related by a linear transformation:
\be\label{duality}
	| \psi_{c,\alpha}(\{h\})\rangle 
	= \sum_{c',\alpha'} u^\sigma_{c,c',\alpha,\alpha'}(\{h\}) \overline{| \psi_{c',\alpha'}(\{h\})\rangle}. 
\ee
Here,
$\sigma$ labels different ``configurations'' of the quasiholes, i.e., 
components of the quasihole configuration space that can be connected
without violating the conditions  $\abs{h_{jx}-h_{ix}}\gg 1$, $\abs{h_{jy}-h_{iy}}\gg 1$.
We define these configuration-labels  with respect to quasihole configurations in the infinite plane.
Toroidal periodicity leads to an equivalence between various triples
$(\sigma, (c,\alpha), (c',\alpha'))$ that will be taken into account later.
The transformation described by $u^\sigma_{c,c',\alpha,\alpha'}(\{h\})$
must also be unitary, since states in different topological sectors are
generally orthogonal.

A lot is known about the transition functions $u^\sigma_{c,c',\alpha,\alpha'}(\{h\})$
from the properties of the zero-mode basis 
states---$|a_1,\dots,a_n;c,\alpha\rangle$ and $\overline{|a_1,\dots,a_n;c,\alpha\rangle}$---under 
magnetic translations alone.\cite{flavin11}
Their dependence on $h$, $(c,c')$, and $(\alpha,\alpha')$ separates into the following factorized form:
\be \label{separate}
	u^\sigma_{c,c',\alpha,\alpha'}(\{h\})=u(\{h\})M_{c,c'}\xi^\sigma_{\alpha,\alpha'},
\ee
with $u(\{h\})$ and $M_{c,c'}$ fully determined by translational symmetry:
\be
	u(\{h\})=\exp \left( \frac{i\pi}3 \sum_j h_{jx}h_{jy} \right),
\ee
\be
	M = \frac 1{\sqrt3} 
	\left( \begin{array}{ccc}
		 e^{2\pi i(L-1)/3} & e^{-2\pi i L/3} & e^{2\pi i/3} \\
		 e^{-2\pi iL/3} & 1 & e^{2\pi iL/3} \\
		 e^{2\pi i/3} & e^{2\pi iL/3} & e^{2\pi i(L-1)/3} 
	\end{array} \right).
\ee

The $h$-independent product $M_{c,c'}\xi^\sigma_{\alpha,\alpha'}$ may be identified as the topological $S$-matrix
of the problem. We will now summarize how the missing information about $\xi^\sigma_{\alpha,\alpha'}$ can be obtained
within this formalism, and subsequently this information can be used to gain knowledge about braiding. For details
we refer the reader to Ref. \onlinecite{flavin11}.

\begin{figure}  
	\includegraphics[width=.9\columnwidth]{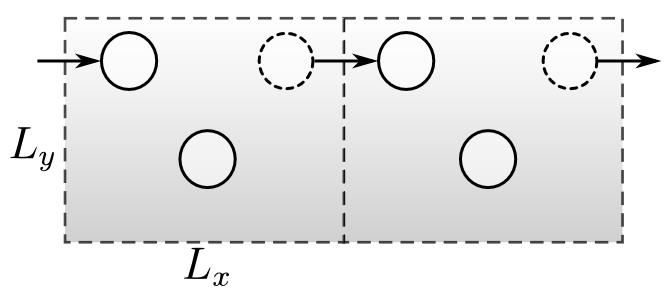} 
	\caption{Two quasiholes, one stationary and one moving across the boundary, on a torus 
		which is shown in a ``repeated zone scheme''. 
		The stationary quasihole is at the bottom of the figure. 
		Initially the moving quasihole (which starts at the dotted-line circle) is ordered second, 
		and the system is in some configuration $\sigma$.
		After that quasihole moves across the boundary it becomes the first, 
		and the system is in configuration $g_x(\sigma)$.}
	\label{xpathfig}
 \end{figure}

Further constraints on the matrix $\xi^\sigma_{\alpha,\alpha'}$ can be obtained by moving (via adiabatic transport) 
the positions of some of the $n$ quasiholes across the boundaries of a rectangular coordinate chart on the torus, which we fix once and for all (see Fig. \ref{xpathfig}). 
It is important to note that this can be done while 
maintaining $\abs{h_{jx}-h_{ix}}\gg 1$ and $\abs{h_{jy}-h_{iy}}\gg 1$, and this leads to new consistency
conditions on the $\xi^\sigma_{\alpha,\alpha'}$ matrix. 
\begin{figure}  
	\includegraphics[width=.55\columnwidth]{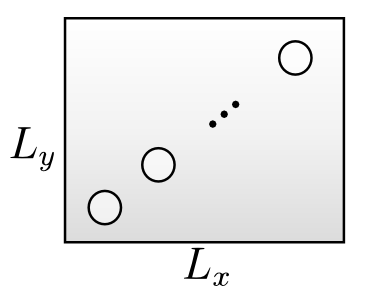} 
	\caption{The configuration $\sigma_I$ is the one in which the leftmost (first) quasihole
		is bottommost, the next to the right (second) is the next above, and so on.}
	\label{sigmaI}
\end{figure}
We begin in a configuration $\sigma$---e.g., the one
shown in Fig. \ref{sigmaI}
---and carry out this process in two steps.
First, we consider the adiabatic transport of the upper right particle along a path that moves it further to the right.
As soon as the particle crosses the right boundary of our coordinate chart,
the particle reappears on the left boundary. Viewed in this coordinate chart, the quasihole configuration 
has changed from $\sigma$ to a new configuration $g_x(\sigma)$. At the same time, with the particle that was
formally rightmost now being leftmost, the topological sector label (the sequence of patterns appearing in the
$L_y\rightarrow 0$ limit) of the state has changed. If $\alpha$ was the label of the topological (super-)sector
initially, the new label $F(\alpha)$ is easily found from the associated thin torus patterns.\cite{flavin11}
Thus, while the transition function expressing the duality relation \eqref{duality} is defined
in terms of $\xi^\sigma_{\alpha,\alpha'}$ right before crossing the boundary of the coordinate chart,
it is defined in terms of $\xi^{g_x(\sigma)}_{F(\alpha),\alpha'}$ right thereafter. Is is easy to work out
a condition enforcing continuity across the boundary, relating these two matrices,\cite{flavin11}
\begin{align} \label{xpath}
	\xi^{g_x(\sigma)}  
	&= B^{-1} \textrm{diag}[ e^{-2\pi i L /3 -i L \delta(\alpha,n)/3-i/3\sum_k\delta(\alpha,k)} ]
	\xi^\sigma \nonumber\\
	& \quad \times \textrm{diag} [ e^{ -2\pi i f_{\sigma_n}(\alpha)/3} ] ,
\end{align}
where the argument of $\mathrm{diag}[\dots]$ specifies the $\alpha$-th diagonal entry of a diagonal matrix. 
The matrix $B$ is defined as $B_{\alpha,\alpha'} = \delta_{\alpha,F(\alpha')}$. 
A similar process can be considered where the topmost quasihole of a configuration $\sigma$ moves up, crossing the
upper boundary of the coordinate chart, becoming the bottommost quasihole in the resulting new configuration
$g_y(\sigma)$. The analogous continuity condition on the $\xi$ matrices reads,
\begin{align} \label{ypath}
	\xi^{g_y(\sigma)} 
	&= \textrm{diag} [ e^{ -2\pi i f_j(\alpha)/3} ]
	\xi^\sigma \nonumber\\
	& \quad \times 
	\textrm{diag}[ e^{-2\pi i L/3-i L \delta(\alpha,n)/3-i/3\sum_k\delta(\alpha,k)} ] 
	B .
\end{align}
The purpose of Eqs. \eqref{xpath} and \eqref{ypath} is twofold: They connect $\xi$ matrices for different quasihole
configurations, which will be essential in describing braiding processes, and they impose constraints on any given
$\xi$ matrix. To see the latter, focus on the quasihole configuration shown in Fig. \ref{sigmaI}, which 
we will now refer to as $\sigma_I$. It is easy to see that $\sigma_I$ is invariant under the two moves 
described above performed in succession, i.e., $g_y(g_x(\sigma_I))=\sigma_I$. Hence,  Eqs. \eqref{xpath} and \eqref{ypath} 
together constrain the matrix elements of $\xi^{\sigma_I}$.

Yet another way in which the transition functions and hence the $\xi$-matrices are related for different
quasihole configurations is by mirror symmetry. In the presence of a constant magnetic background 
field, mirror symmetries exist only in conjunction with time-reversal symmetry, which we will leave understood.
The operators $\tau_x$ and $\tau_y$ associated with mirror reflections across the $x$ and $y$ axis are therefore
anti-linear operators. It is then simple to find relations between the $\xi$-matrices using these operations,\cite{flavin11}
\begin{align}
	\label{taux}
	\xi^{g_{\tau_x}(\sigma)} 
		&= (B_\tau)^{-1} 
		\textrm{diag} [ e^{i L/3\sum_k \delta(\alpha,k)} ]
		(\xi^\sigma)^* \nonumber \\
		&\quad \times \textrm{diag} [e^{2\pi i/3 \sum_k (1+\delta(\alpha,k)/\pi)f_k(\alpha)} ] ,\\
	\label{tauy}
	\xi^{g_{\tau_y}(\sigma)} 
		&=
		\textrm{diag} [e^{2\pi i/3 \sum_k (1+\delta(\alpha,k)/\pi)f_k(\alpha)} ]
		(\xi^\sigma)^* \nonumber\\
		&\quad \times \textrm{diag} [e^{i L/3\sum_k \delta(\alpha,k)} ]
		B_\tau .
\end{align}
In the above, $g_{\tau_x}(\sigma)$ is the configuration that is the mirror image of $\sigma$ under $\tau_x$,
and similarly $g_{\tau_y}(\sigma)$.
The matrix $B_\tau$ appearing above is defined by $(B_\tau)_{\alpha,\alpha'} = \delta_{\alpha,F_\tau(\alpha')}$, where $F_\tau(\alpha)$ is the 
(super-)sector resulting from an inversion of the patterns associated with the (super-)sector $\alpha$.
Operating $\tau_x$ and $\tau_y$ in succession on a system in the configuration $\sigma_I$ gives $\sigma_I$ again, i.e., $g_{\tau_y}(g_{\tau_x}(\sigma_I))=\sigma_I$. 
In this way Eqs. \eqref{taux} and \eqref{tauy} allow us to further constrain the elements of $\xi^{\sigma_I}$. 

\begin{figure}  
	\includegraphics[width=.75\columnwidth]{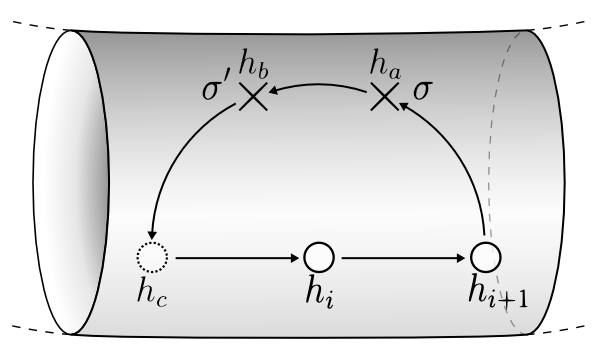} 
	\caption{Exchange path for two quasiholes at positions $h_i$ and $h_{i+1}$. 
		First, the quasihole at $h_{i+1}$ is dragged to $h_a$. There the coherent state representation is 
		changed from the original basis to the dual basis using $\Xi^\sigma$. That quasihole 
		is then dragged to $h_b$, and the state is changed back to the original basis using 
		$\Xi^{\sigma'}$. The quasihole at $h_b$, now the $i$-th, is moved to $h_c$, then both quasiholes 
		are moved to their final positions: The quasihole at 
		$h_i$ goes to $h_{i+1}$ and the quasihole at $h_c$ goes to $h_i$.
		Using the coherent state Ansatz, this gives rise to \Eq{nonab}.
		}
	\label{braidingpath}
 \end{figure}
 
To braid two quasiholes, we adiabatically transport them along the path in \Fig{braidingpath}. 
To this end, we consider a configuration $\sigma$ where both quasiholes are adjacent both
in terms of their $x$ and their $y$ coordinates, i.e., they can be exchanged without violating
$\abs{h_{jx}-h_{ix}}\gg 1$ or $\abs{h_{jy}-h_{iy}}\gg 1$ for any {\em other} pair of
quasiholes. The result of the adiabatic transport can then be worked out 
directly from the coherent state Ansatz \eqref{psi} given known transition functions, i.e.,
the matrices $\xi^\sigma$.\cite{seidel_lee, seidel_pfaffian, flavin11} It can be expressed as follows:
\be\label{nonab}
	\left( \begin{array}{c}
		| \Psi_1 \rangle \\ | \Psi_2 \rangle \\ \vdots \\ | \Psi_n \rangle
	\end{array} \right)
	\rightarrow e^{i \Phi_{\text{AB}}}
	\Xi^\sigma (\Xi^{\sigma'})^\dagger
	\left( \begin{array}{c}
		| \Psi_1 \rangle \\ | \Psi_2 \rangle \\ \vdots \\ | \Psi_n \rangle
	\end{array} \right)\,.
\ee
Here, $\Phi_{\text{AB}}$ is the Aharonov-Bohm phase, equal to the charge of a quasihole, $-1/3$, 
times the area enclosed by the braiding path.
The quantities $|\Psi_\alpha\rangle$ are the three-component column vectors with entires 
$|\psi_{c,\alpha}\rangle$. $\Xi^\sigma$ is the
matrix $\xi^\sigma\otimes M$ (where we identify the states $|\psi_{c,\alpha}\rangle$ 
with a formal tensor product basis $|\alpha\rangle\otimes |c\rangle$). 
$\sigma$ is the initial configuration of the quasiholes, and $\sigma'$ the other 
configuration that occurs during braiding (Fig. \ref{braidingpath}),
i.e., the one obtained from $\sigma$ by crossing the line $h_{jx}=h_{ix}$ or the line $h_{jy}=h_{iy}$.

It turns out that the result of braiding is always block diagonal in the $c$ labels, i.e., the braid matrix
$\Xi^\sigma (\Xi^{\sigma'})^\dagger$ is of the form $\chi_i(n)\otimes\mathbb I_{c_\text{max}\times c_\text{max}}$,
where we call $\chi_i(n)=\xi^\sigma (\xi^{\sigma'})^\dagger$ the ``reduced'' braid matrix associated with
a counter-clockwise exchange 
of the $i$-th and $i+1$-st of $n$ quasiholes.
This fact is a direct consequence of translational invariance. Moreover,  $\chi_i(n)$  is found to be
independent of the initial configuration $\sigma$, as one would expect.

The various constraint equations discussed so far still admit many solutions for the braid matrix.
This is chiefly due to the fact that the asymmetry parameter $s$ introduced above is still
undefined. A final set of constraint equations comes from the imposition of certain locality constraints
on the braid matrix that can be motivated directly from the thin torus patterns.\cite{seidel_pfaffian, flavin11}
Said succinctly, we mean by locality that the result of braiding should only depend on the
sequence of three ground-state patterns forming the two domain walls associated with the braided quasiholes, and
that only the pattern in the middle may change as a result of braiding.
The constraint equations following from this, combined with the above symmetries, then 
lead to a discrete set of (usually intimately related) solutions for the statistics.
We will discuss the full set of constraint equations and their solutions in App. \ref{app2} for the two-quasihole
case, and in App. \ref{app3} for the three-quasihole case.
Here we will summarize the results of this calculation by giving the (reduced) braid matrices obtained from it for 
both two and three particles:
\begin{align} 
	\chi_1(2) &= \xi^{(1,2)}(\xi^{(2,1)})^\dagger \nonumber \\
	&= e^{-i \pi /3}
	\left( \begin{array}{ccc}
		e^{-i\pi a} && \\
		& e^{i \pi a} \varphi^{-1} & e^{-2i \pi a} \varphi^{-1/2} \\
		& e^{-2i \pi a} \varphi^{-1/2} & \varphi^{-1}
	\end{array} \right),
	\label{chi2}
\end{align}
\begin{align}
	\chi_1(3) &= \xi^{(1,2,3)}(\xi^{(2,1,3)})^\dagger \nonumber\\
	&= e^{-i \pi /3}
	\left( \begin{array}{cccc}
		e^{2i \pi a} &&& \\
		& e^{2i\pi a} && \\
		&& e^{i \pi a} \varphi^{-1} & e^{-2i\pi a} \varphi^{-1/2} \\
		&& e^{-2i\pi a} \varphi^{-1/2} & \varphi^{-1}
	\end{array} \right),
	\label{chi3}
\end{align}
where $a=\pm 1/5$, and $\varphi$ is the golden ratio, $\varphi=(1+\sqrt 5)/2$.
Here, the rows and columns refer to the $\alpha$-supersectors of Tables \ref{table_2sectors} and \ref{table_3sectors}.
The parameter $s$ is found to have one of two values: $s=2-3a/2$.
Just as in the case of the $k=3$ Read-Rezayi state discussed in Ref. \onlinecite{flavin11},
there are two solutions, one for each sign of $a$, 
which are related by an Abelian phase and complex conjugation.
In the above, we have also fixed a gauge degree of freedom associated with unitary transformations.
\begin{table}[!tbp]  
   \begin{tabular}{ccc | c || c | c | c }
	& $\alpha$ && Thin torus pattern & $f_1(\alpha)$ & $f_2(\alpha)$ & $f_3(\alpha)$ \\
	\hline
	& 1 && $002\mathbf{00\underline{1}}1011011\mathbf{0\underline{1}0}1101101\mathbf{\underline{1}00}2002$ 
		& $-s$ & 0 & $s$ \\
	& 2 && $11011\mathbf{\underline{0}10}1101\mathbf{10\underline{0}}2002002\mathbf{0\underline{0}1}10110$ 
		& 1 & $-2+s$ & $1-s$ \\
	& 3 && $1101\mathbf{1\underline{0}0}2002002\mathbf{\underline{0}01}1011\mathbf{01\underline{0}}110110$ 
		& $-1+s$ & $2-s$ & $-1$ \\
	& 4 && $11011\mathbf{\underline{0}10}11011\mathbf{0\underline{1}0}11011\mathbf{01\underline{0}}110110$ 
		& $1$ & $0$ & $-1$
   \end{tabular} 
   \caption{$c=0$ thin torus patterns for a three-quasihole Gaffnian state, 
   	and the offset functions of the associated domain walls. The elementary domain wall strings are in 
	bold, and the orbital positions, $3n_j$, are underlined. 
   	Patterns for $c=1(-1)$ can be obtained by shifting each occupancy number one orbital to the
	right (left), and the shift functions obtained using $f_j(c,\alpha)=f_j(\alpha)+c$.} 
   \label{table_3sectors} 
\end{table}

Together with the locality constraint described above, these two matrices imply
the result of braiding any adjacent pair in a state of $n$ quasiholes.
A ``tensor representation'' of the statistics just as discussed in Ref. \onlinecite{slingerland}
can then immediately be constructed in complete analogy with Ref. \onlinecite{flavin11}.

\section{\label{discussion}Discussion}
\begin{figure}  
	\includegraphics[width=.75\columnwidth]{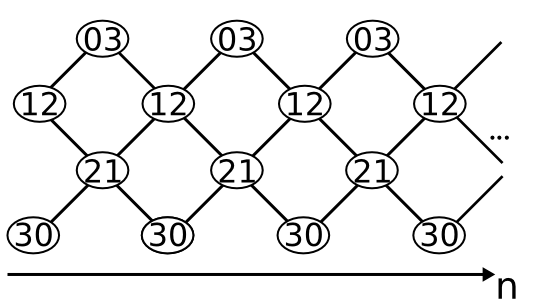} 
	\includegraphics[width=.75\columnwidth]{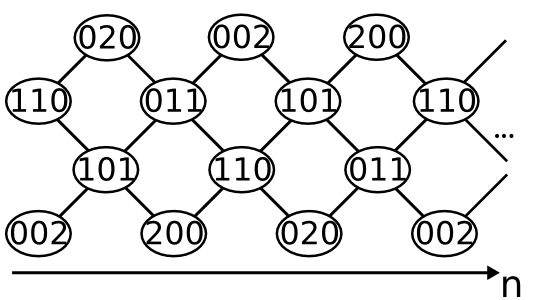}
	\caption{Bratteli diagrams depicting the $c=0$ patterns in the RR state (top) and the Gaffnian (bottom).
		Valid topological sectors on the torus 
		are represented by paths which start at the left, take one step right 
		(either up or down) for each quasihole in the state,
		and end on the same type of ground state 
		pattern---(200)-type or (011)-type---as they began, minding periodic boundary conditions.
		There is a one-to-one correspondence between the paths in the lower diagram and the
		patterns in Tables \ref{table_2sectors} and \ref{table_3sectors}, and also between the valid paths in the
		upper and lower diagrams. This latter correspondence is how the SU$(2)_3$ and SU$(3)_2$
		rank-level duality manifests in terms of patterns. 
		It should be noted that for each sector which corresponds to a path in one of these diagrams,
		there is for RR an additional sector related to the first by translation, 
		and two additional translated sectors for the Gaffnian.
		}
	\label{bratteli}
 \end{figure}

The two solutions obtained in the preceding Sec. \ref{braid} 
are related to one another simply by complex conjugation and an overall Abelian phase.
They describe Fibonacci anyons and are thus closely related
to those obtained for the $k=3$ Read-Rezayi (RR) state using the same method.\cite{flavin11}
In essence, the solutions obtained from the RR patterns and those obtained here are
the same up to an Abelian phase. This statement excludes ``global exchange paths'' on the
torus involving processes such as the one depicted in \Fig{xpathfig}, 
as we will further explain below.
This close correspondence between the solutions found from RR and Gaffnian patterns
is a manifestation of level-rank duality between the associated SU$(2)_3$ and SU$(3)_2$ 
fusion rules, respectively. This duality was also discussed by Ardonne\cite{ardonne_gaffnian} in terms of 
domain walls. It is manifest in the Bratteli diagrams of \Fig{bratteli}, which in the present
context represent the rules for domain wall formation between ground state patterns for the 
respective states. Note, however, that there is a three to two correspondence between the topological
sectors in both cases, rather than one-to-one. This is so since each sector, i.e., each path in the
Bratteli diagram, is threefold degenerate under translations in the Gaffnian case, but only twofold 
in the $k=3$ RR case (at filling factor $3/2$). The correspondence between Gaffnian and RR
sectors is perfect if we limit ourselves to the $n-1$ generators of the braid group $\sigma_{i,i+1}$, $i=1,\dotsc, n-1$,
which exchange the $i$-th and $i+1$-st quasihole. These generate the full braid group in the plane, but not
on the torus.  On the torus, these generators leave certain subspaces of topological sectors invariant, which all
start and end in the same pattern in the topological sector label. These subspaces for the Gaffnian are then in
correspondence with similar subspaces for the RR state, in the sense that there are isomorphisms between them that
commute with braiding, except for an overall Abelian phase. This correspondence, however, gets spoiled 
by the inclusion of the remaining generators on the torus, which mix the subspaces. This happens differently
for the Gaffnian and the RR case, since there are six such subsectors in the former case, but only four in the latter.

The subtle differences between our solutions for the Gaffnian and the RR case on the torus are of a piece
with the difference in overall Abelian phase. It is well known that the overall Abelian phase could in principle
assume any value in planar geometry, but on the torus, it is constrained by the topological degeneracies 
characterizing the state. In the coherent state method, one source of phase differences is the factor
$i/3$ in the coherent state Ansatz, \Eq{phi}, which is generally related to the ``stride'' of the domain wall
in a given topologic sector, which equals $3$ in the present case and $2$ in the RR case. This stride is of course identical to the center-of-mass degeneracy. In particular, one may see that the equations obtained from global processes 
such as the one shown in \Fig{xpathfig} 
are quite sensitive to this stride and the associated phase (see \Eq{app2_globalpaths} below). 
In view of the importance
of these processes in our method, and the fact that they spoil the correspondence between Gaffnian and RR 
topological sectors as explained above, it may not be clear {\em a priori} that the consistency equations
we obtain in both cases admit closely related solutions, in the sense discussed.
That this is so can be traced back to the fact that the translational degree of freedom, $c$, decouples early on 
(see below \Eq{separate}), and the remaining $\alpha$ degree of freedom is fully analogous in both cases.
This is how rank-level duality becomes manifest in the present formalism. For similar reasons, our
solutions for the Gaffnian and RR states, which were both obtained at the maximum (bosonic) filling factor, could be 
generalized quite easily to lower filling factors.

As emphasized initially, the Fibonacci-type solutions we obtained are distinct from the anyon model
associated with the conformal block monodromies of the Gaffnian state. The latter describes
so-called Yang-Lee anyons, whose relation to Fibonacci anyons and the associated Galois duality
has enjoyed much interest recently.\cite{ardonne_yanglee, freedman_galois} 
Yang-Lee anyons are associated with nonunitary $F$-matrices consistent with the SU$(3)_2$ fusion 
rules. It has been known for some time, however, that the same fusion rules admit unitary solutions 
of the Fibonacci type also; these are realized by the NASS state of Ref. \onlinecite{ardonne3}. 
Indeed, it is not difficult to perform checks confirming that one of our solutions, that corresponding to $s=17/10$, 
agrees exactly with the monodromies of the NASS state (including the overall phase). The dominance patterns 
of the NASS state have been discussed more recently,\cite{ardonne_regnault} and it seems clear that the 
calculation presented here can be carried over to this state without essential changes. 
In all, this confirms once more for the 
case of SU$(3)_2$ that the coherent state Ansatz  produces a subset of all {\em unitary} anyon models consistent with
given fusion rules.

In the case of the unitary NASS state, our results support the usual conjecture that the holonomies associated
with adiabatic transport agree with conformal block monodromies. In the case of the Gaffnian, things are more subtle.
We have argued that in this case, provided that the coherent state Ansatz is justified, the holonomies gives rise 
to a well defined Fibonacci anyon statistics, possibly (given the twofold ambiguity of our solution) identical to those of the NASS state. In particular, we believe that our Ansatz
is indeed well justified if the gap of the Gaffnian parent Hamiltonian does not close in the charge sector.
If so, this would presumably remain a formal property of the Gaffnian state that is not robust to perturbations,
even ones that do not open a gap in the neutral sector. However, it might shine new light on the formal connections between the Gaffnian
and the NASS state, which have been discussed previously.\cite{ardonne_gaffnian}
We leave more rigorous investigations into this matter for future studies.

\section{Conclusion}
In this paper, we have analyzed the effect of exchanging quasiholes described by Gaffnian quantum Hall trial state wave functions via adiabatic transport, based on a coherent
state Ansatz in terms of adiabatically continued thin-torus zero modes. We have argued that this Ansatz may correctly describe the holonomies of Gaffnian quasiholes
as long as the charge sector of the state remains gapped. In this case, we find that the statistics obtained are either closely related, or in fact identical, to those of the
NASS state, though this property will not be topologically protected. This observation is of a kind with the statement that these states have the same fusion rules,
though their conformal block monodromies are different,
and are non-unitary in the Gaffnian case. 
Our results appear to confirm in yet another case that the coherent state Ansatz produces a set of unitary anyon
models consistent with the fusion rules of the underlying state.

\begin{acknowledgments}
JF and AS are indebted to S. Simon for helpful discussions on Gaffnian monodromies.
RT thanks N. Regnault for discussions as well as B. A. Bernevig and
M. Greiter for collaborations on related topics.
This work was supported  by the National Science Foundation under NSF
Grant No. DMR-0907793. RT is supported by an SITP fellowship by
Stanford University.
\end{acknowledgments}

\appendix
\section{\label{app2}Two-quasihole Solution}
To work out the braid group for $n$ quasiholes with general $n$, one needs to consider only 
braiding for pairs of quasiholes associated with all possible pairs of domain walls,
as given by all possible combinations of three ground state patterns.
Locality then implies that all the other ground state patterns appearing in the topological
sector label will not affect the result of braiding. 
To this end, we will solve for the reduced braid matrix in the simple cases of $n=2$ (here) and
$n=3$ (in Appendix \ref{app3}).
Together, these results can be used to construct braid matrices
for $n$-quasihole states,
since these cases exhaust all possible sequences of three ground state patterns
separated by domain walls.


The set of constraint equations to solve for the reduced braid matrix come from enforcing the unitarity of the $\xi$ matrices, the locality condition discussed in the main
text, and the global path relations  Eqs. \eqref{xpath} and \eqref{ypath}. We will begin by enforcing the latter.
As discussed above, we can apply Eqs. \eqref{xpath} and \eqref{ypath} in succession to constrain $\xi^{\sigma_I}$ because $\sigma_I=g_y(g_x(\sigma_I))$.
Applying these two equations, with the data from Table \ref{table_2sectors}, results in the constraint equation
\be \label{app2_globalpaths}
	\xi^{\sigma_I} =
	\left( \begin{array}{ccc}
		& \Delta p^{-1} & \\
		\Delta p & 0 & \\
		&& -1
	\end{array} \right)
	\xi^{\sigma_I}
	\left( \begin{array}{ccc}
		& \Delta p & \\
		\Delta p^{-1} & 0 & \\
		&& -1
	\end{array} \right) ,
\ee
where $p=-\exp\left[-2\pi i(1+s)/3\right]$, $\Delta = \exp\left[-2\pi i(L/2+1)D/3 \right]$, and $D = 0$ or 1 if the $\delta$ parameter for the $100$-type domain walls is 0 or $\pi$, respectively. Equation \eqref{app2_globalpaths} is satisfied when
\be \label{app2_xi}
	\xi^{\sigma_I} =
	\left( \begin{array}{ccc}
		\xi_{11} & \xi_{12} & \xi_{13} \\
		\xi_{12} & p^2 \xi_{11} & -\Delta p \xi_{13} \\
		\xi_{31} & -\Delta p \xi_{31} & \xi_{33}
	\end{array} \right) .
\ee
Mirror symmetry, Eqs. \eqref{taux} and \eqref{tauy}, can also produce a constraint equation because $g_{\tau_y}(g_{\tau_x}(\sigma_I))=\sigma_I$. However, in the case of two quasiholes, applying Eqs. \eqref{taux} and \eqref{tauy} in succession results in the trivial equation, $\xi^{\sigma_I} = \xi^{\sigma_I}$.

The process of solving for the reduced braid matrix is similar to the solution 
given for $n=2$
in Ref. \onlinecite{flavin11}. We gain the following equations by demanding that $\xi^{\sigma_I}$ is unitary, 
\bsub \label{app2_unitary} \begin{align}
	e^{i\pi/3} &= 2 \Delta \eta^{-D}p \xi_{11}\xi_{12}+\eta^D p \xi_{13}^{\phantom{13}2}, \\
	e^{i\pi/3} &= 2 \eta^Dp \xi_{31}^{\phantom{31}2}-\xi_{33}^{\phantom{33}2}, \\ 
	0 &= \Delta \eta^{-D}p^3\xi_{11}^{\phantom{11}2} + 
		\Delta\eta^{-D}p\xi_{12}^{\phantom{12}2} - \Delta\eta^Dp^2\xi_{13}^{\phantom{13}2}, \\
	0 &= \eta^{-D}\xi_{31} \left(-p^2\xi_{11}+\Delta p\xi_{12} \right) + \eta^D p \xi_{13}\xi_{33},
\end{align} \esub 
where $\eta = \exp\left( -2\pi i/3 \right)$.
Two additional equations come from the requirement that braiding is local; 
as said above, this means that the result of braiding should only depend on the
sequence of three ground-state patterns forming the two domain walls associated with the braided quasiholes, and
that only the pattern in the middle may change as a result of braiding. 
Imposing these locality considerations tells us that $\chi_1(2)$ must be of the form
\be \label{app2_chilocal}
	\chi_1(2) = 
	\left( \begin{array}{ccc}
		\cdot & 0 & 0 \\
		0 & \cdot & \cdot \\
		0 & \cdot & \cdot
	\end{array} \right),
\ee
where ``$\cdot$''s are unknown, potentially nonzero, matrix elements for which we will solve.
By applying the form in \Eq{app2_chilocal} to the matrix $\chi_1(2)=\xi^{\sigma_I}(\xi^{{\sigma_I}'})^\dagger$ derived from adiabatic transport, 
the zero elements give two more independent constraint equations, 
\bsub \label{app2_local} \begin{align}
	0 &= \eta^{-D}(1+p^2)\xi_{11}\xi_{12}-\Delta\eta^D\xi_{13}^{\phantom{13}2}, \\
	0 &= \eta^{-D}\xi_{31} \left( \xi_{11}-\Delta p\xi_{12} \right) + \eta^D\xi_{13}\xi_{33}.
\end{align} \esub

Solving this system of six equations, \eqref{app2_unitary} and \eqref{app2_local}, 
is formally similar to the solution in the Appendices of Ref. \onlinecite{flavin11}, so the details will not be repeated here. 
Just as in that reference, there are two solutions: A special solution in which $p=\pm i$, 
\bsub \label{app2_special} \begin{align}
	\xi_{11}^{\phantom{11}2} &= \frac{1}{2}\eta^D e^{i\pi/3}p^{-1}, \\
	\xi_{12} &= \Delta \xi_{11}, \\
	\xi_{13} &= \xi_{31} = 0, \\
	\xi_{33}^{\phantom{33}2} &= -e^{i\pi/3},
\end{align} \esub
which produces the reduced braid matrix
\be \label{app2_chispecial}
	\chi_1(2) = e^{2\pi i/3}
	\left( \begin{array}{ccc}
		p & 0 & 0 \\
		0 & p & 0 \\
		0 & 0 & e^{-i\pi/3}
	\end{array} \right)
\ee
(but we will show in Appendix \ref{app3} that this solution is inconsistent with the equations from three-quasihole braiding), and the consistent solution,
\bsub \begin{align}
	\xi_{11}^{\phantom{11}2} &= \frac{\eta^D e^{i\pi/3}}{(1+p)^2}, \\
	\xi_{12} &= \Delta \xi_{11}, \\
	\xi_{13}^{\phantom{13}2} &= \eta^D (p+p^{-1})\xi_{11}^{\phantom{11}2}, \\
	\xi_{31}^{\phantom{31}2} &= \eta^D (p+p^{-1})\xi_{11}^{\phantom{11}2}, \\
	\xi_{33}^{\phantom{33}2} &= \eta^{-D} (1-p)^2 \xi_{11}^{\phantom{11}2},
\end{align} \esub
which produces the reduced braid matrix
\begin{align} \label{app2_chi}
	&\chi_1(2) = e^{-i\pi/3} \times \nonumber \\
	& \left( \begin{array}{ccc}
		p^{-1} & 0 & 0 \\
		0 & p(p+p^{-1}-1) & \pm e^{i \pi D /3}(1-p) \sqrt{p+p^{-1}}\\
		0 & \pm e^{-i \pi D /3}(1-p) \sqrt{p+p^{-1}} & p+p^{-1}-1
	\end{array} \right).
\end{align}
This two-quasihole reduced braid matrix is the same as that in Ref. \onlinecite{flavin11} except for two features:
The $D$-dependent phase on the off-diagonal elements 
(though this will later be removed with a unitary transformation), 
and the overall Abelian phase, which here is $e^{-i\pi/3}$ and in Ref. \onlinecite{flavin11} was $e^{i\pi/2}$.

\section{\label{app3}Three-quasihole Solution}
The solution for the reduced braid matrix of three quasiholes begins similarly to that for two quasiholes. 
We first constrain $\xi^{\sigma_I}$ using global path relations, Eqs. \eqref{xpath} and \eqref{ypath}, 
and mirror symmetry, Eqs. \eqref{taux} and \eqref{tauy}.
Applying the former two in succession and filling in the data from Table \ref{table_3sectors} gives the constraint
\be \label{app3_globalpaths}
	\xi^{\sigma_I} = 
	\left( \begin{array}{cccc}
		 & \tilde\Delta p^{-1} &  &  \\
		 & 0 & -\eta^D &  \\
		\tilde\Delta p &  &  &  \\
		 &  &  & -1 \\
	\end{array} \right)
	\xi^{\sigma_I}
	\left( \begin{array}{cccc}
		 &  & \tilde\Delta p &  \\
		\tilde\Delta p^{-1} & 0 &  &  \\
		 & -\eta^D &  &  \\
		 &  &  & -1 \\
	\end{array} \right) ,
\ee
where $p$ is defined as in Appendix \ref{app2}, and $\tilde\Delta = \exp\left[-2\pi i (L/2+1)D/3 \right]$. 
This definition for $\tilde\Delta$ is seemingly the same as that for $\Delta$ in Appendix \ref{app2}, 
but in the case of two quasiholes $L=1$ modulo 3, and here $L=0$ modulo 3.

The two mirror symmetry equations, \eqref{taux} and \eqref{tauy}, can be applied in succession to $\xi^{\sigma_I}$ to give
\be \label{app3_mirror}
	\xi^{\sigma_I} = 
	\left( \begin{array}{cccc}
		1 &  &  &  \\
		 & 0 & \eta^D &  \\
		 & \eta^{-D} & 0 &  \\
		 &  &  & 1 
	\end{array} \right)
	\xi^{\sigma_I}
	\left( \begin{array}{cccc}
		1 &  &  &  \\
		 & 0 & \eta^{-D} &  \\
		 & \eta^D & 0 &  \\
		 &  &  & 1 
	\end{array} \right) .
\ee
Equations \eqref{app3_globalpaths} and \eqref{app3_mirror} together constrain $\xi^{\sigma_I}$ to be of the form
\be \label{app3_xi}
	\xi^{\sigma_I} =
	\left( \begin{array}{cccc}
		\xi_{11} & \xi_{12} & \eta^{-D} \xi_{12} & \xi_{14} \\
		\xi_{12} & \eta^D p^2 \xi_{11} & -\tilde\Delta p \xi_{12} & -\tilde\Delta^{-1} p \xi_{14} \\
		\eta^{-D} \xi_{12} & \tilde\Delta p \xi_{12} & \eta^{-D} p^2 \xi_{11} & -\tilde\Delta p \xi_{14} \\
		\xi_{41} & -\tilde\Delta^{-1} p \xi_{41} & -\tilde\Delta p \xi_{41} & \xi_{44}
	\end{array} \right) .
\ee
The structure of this matrix is almost identical to the corresponding matrix $\xi^{++}$ in Ref. \onlinecite{flavin11}, save that the $D$-dependent phases $\eta^D$ and $\tilde\Delta$ are different.
There, $\tilde\Delta$ was defined such that $\tilde\Delta^2 = e^{i\pi D}$, 
whereas here $\tilde\Delta^2 = e^{2i\pi D/3}=\eta^{-D}$.
We might then find a braid matrix with a non-trivial dependence on $D$.
However, the symmetry relations in Eqs. \eqref{xpath}, \eqref{ypath}, \eqref{taux}, and \eqref{tauy} 
also have additional $D$-dependent phases compared to their corresponding forms in Ref. \onlinecite{flavin11},
and we will see that by following the same steps as in that reference to find the braid matrix solution,
all the $D$-dependent phases will conspire to cancel save for those on the off-diagonal elements
which can be removed via a unitary transformation.

By requiring that $\xi^{\sigma_I}$ be unitary, we find the following constraint equations,
\bsub \label{app3_unitary} \begin{align}
	1 &= \abs{\xi_{11}}^2+2\abs{\xi_{12}}^2+\abs{\xi_{14}}^2, \\
	0 &= \tilde\Delta \xi_{12} \xi_{11}^{\phantom{11}*} + \tilde\Delta^{-1} p^2 \xi_{11}\xi_{12}^{\phantom{12}*}
		-p \abs{\xi_{12}}^2 -p\abs{\xi_{14}}^2, \\
	0 &= \xi_{41} \left( \xi_{11}^{\phantom{11}*} - 2\tilde\Delta^{-1} p\xi_{12}^{\phantom{12}*} \right)
		+\xi_{14}^{\phantom{14}*}\xi_{44}, \\
	1 &= 3\abs{\xi_{41}}^2 + \abs{\xi_{44}}^2.
\end{align}\esub
The locality of braiding tells us not only that some elements of $\chi_1(3)$ must be zero, 
as was the case for two quasiholes, 
but also that the $2\times2$ block with off-diagonal elements 
must be equal to the equivalent $2\times2$ block in $\chi_1(2)$.
This is because the sequences of ground state patterns of the domain walls associated
with the quasiholes to be braided are the same for those two supersectors in
the two- and three-quasihole cases. 
In other words, $\chi_1(3)$ must be of the form
\begin{align} \label{app3_chilocal}
	&\chi_1(3) = e^{-i\pi/3} \times \nonumber\\
	& \left( \begin{array}{cccc}
		\cdot &&&\\
		& \cdot && \\
		&& p(p+p^{-1}+1) & \pm e^{i \pi D /3}(1-p) \sqrt{p+p^{-1}}\\
		&& \pm e^{-i \pi D /3}(1-p) \sqrt{p+p^{-1}} & p+p^{-1}+1
	\end{array} \right),
\end{align}
if \Eq{app2_chi} is the correct reduced braid matrix for two-quasiholes,  
and of the form
\be \label{app3_chilocal2}
	\chi_1(3) = e^{2\pi i/3}
	\left( \begin{array}{cccc}
		\cdot &&&\\
		& \cdot &&\\
		&& p & 0\\
		&& 0 & e^{-i\pi/3}
	\end{array} \right),
\ee
with $p=\pm i$, if \Eq{app2_chispecial} is the correct matrix
(which we will show is not the case).
We find our constraint equations by equating the product 
$\chi_1(3)=\xi^{\sigma_I}(\xi^{{\sigma_I}'})^\dagger$, obtained from adiabatic transport, with the forms above. 
We can perform the two solutions in parallel by using only the elements of 
Eqs. \eqref{app3_chilocal} and \eqref{app3_chilocal2} that are zero in both.
This produces the constraint equations
\bsub \label{app3_local} \begin{align}
	0 &= p^3 \xi_{11}^{\phantom{11}2} -\tilde\Delta^2(p-1)p\xi_{12}^{\phantom{12}2}
		-p^2 \xi_{14}^{\phantom{14}2}, \\
	0 &= \tilde\Delta(p-1)p^2\xi_{11}\xi_{12} +\tilde\Delta^2 p \xi_{12}^{\phantom{12}2}
		-p^2 \xi_{14}^{\phantom{14}2}, \\
	0 &= - \xi_{41} \left[ p\xi_{11}+\tilde\Delta(p-1)\xi_{12} \right]
		+ \xi_{14}\xi_{44}.
\end{align} \esub
The set of constraint equations, \eqref{app3_unitary} and \eqref{app3_local}, is solved when
\bsub \begin{align}
	\xi_{11}^{\phantom{11}2} &= \frac{e^{i\theta_1}}{(1+p)^2}, \\
	\xi_{12} &= \tilde\Delta^{-1}\xi_{11}, \\
	\xi_{14}^{\phantom{14}2} &= (p+p^{-1}-1)\xi_{11}^{\phantom{11}2}, \\
	\xi_{41}^{\phantom{41}2} &= e^{2i\theta_2} (p+p^{-1}-1) \xi_{11}^{\phantom{11}2},\\
\end{align} \esub
which produces the reduced braid matrix
\begin{align} \label{app3_chi}
	&\chi_1(3) = e^{-i\pi/3+i\theta_1} \times \nonumber\\
	& \left( \begin{array}{cccc}
		1 \\ &1 \\
		&& p(1-p) & \pm e^{i\theta_2+i\pi D/3} p \sqrt{p+p^{-1}-1} \\
		&& \pm e^{i\theta_2-i\pi D/3} p \sqrt{p+p^{-1}-1} & e^{2i\theta_2}(1-p)
	\end{array} \right).
\end{align}
Just as in Appendix \ref{app2}, \Eq{app3_chi} is the same reduced braid matrix as was found
in Ref. \onlinecite{flavin11} for three quasiholes, except that here the off-diagonal elements have an
additional $D$-dependent phase and the overall Abelian phase is different.

We have yet to enforce consistency between the $2\times2$ blocks in the two- and three-quasihole
braid matrices; 
to do so we equate \Eq{app3_chi} to Eqs. \eqref{app3_chilocal} and \eqref{app3_chilocal2} in turn.
The latter produces a contradiction, because the off-diagonal elements of \Eq{app3_chi} are not
zero for $p=\pm i$. 
Thus \Eq{app2_special} is not a consistent solution for two quasiholes, and Eqs.
\eqref{app2_chispecial} and \eqref{app3_chilocal2} are not consistent braid matrices.
Enforcing consistency between Eqs. \eqref{app3_chilocal} and \eqref{app3_chi}
implies that
\be
	p+p^{-1} = \varphi,
\ee
where $\varphi$ is the golden ratio, $\varphi=(1+\sqrt 5)/2$.
In other words,
\be
	p = \exp \left( \pm \frac{i \pi}{5} \right),
\ee
or, if we define $a=\pm 1/5$, $p = \exp \left( i \pi a \right)$.
This is all the same as in Ref. \onlinecite{flavin11}. 
However, here the $s$ parameter is defined differently in terms of $p$ than in Ref. \onlinecite{flavin11};
we have defined $p=-\exp\left[-2\pi i(1+s)/3\right]$,
so the $s$ parameter is also constrained to be
\be
	s = 2 - \frac{3 a}{2}.
\ee
Consistency also implies
\begin{align}
	e^{i\theta_1} &= p^2, \\
	e^{2i\theta_2} &= 1.
\end{align}
The expressions for $\chi_1(2)$ and $\chi_1(3)$ in Eqs. \eqref{chi2} and \eqref{chi3}, respectively, 
have been simplified with respect to Eqs. \eqref{app2_chi} and \eqref{app3_chi} and have undergone
a unitary transformation in which each state is multiplied by $(-e^{i\pi D/3})^{\#200}$, 
where $\#200$ is the number of $200200\dots$ strings in the thin torus pattern 
associated with that state. This unitary transformation removes the dependence on the unkonwn
parameter $D$.

\bibliography{library}

\end{document}